\documentclass[aps,prd,twocolumn,floats,nofootinbib,longbibliography,showkeys]{revtex4-1}
\usepackage[dvips]{graphicx}

\usepackage{amsmath}
\usepackage{braket}
\usepackage{amsfonts}
\usepackage[colorlinks=true]{hyperref}
\usepackage{hyperref}
\usepackage{xcolor}

\usepackage{booktabs}

\usepackage{aas_macros}

\usepackage{caption}
\usepackage{graphicx}
\usepackage{subcaption}

\usepackage{bm}
\usepackage{slashed}

\usepackage{changes}

\DeclareMathAlphabet{\mathpzc}{OT1}{pzc}{m}{it}

\def\aap{\textit{Astronomy and Astrophysics}}

\def\be{\begin{equation}}
\def\ee{\end{equation}}
\def\bea{\begin{eqnarray}}
\def\eea{\end{eqnarray}}

\def\L{\mathcal{L}}
\def\Lm{\mathcal{L}_m}

\newcommand\D{\mathcal D}

\begin{document}
\title{Quantum of action in entangled relativity}

\author{Olivier Minazzoli}
\email[]{ominazzoli@gmail.com}
\affiliation{Artemis, Universit\'e C\^ote d'Azur, CNRS, Observatoire C\^ote d'Azur, BP4229, 06304, Nice Cedex 4, France,\\Bureau des Affaires Spatiales, 2 rue du Gabian, 98000  Monaco}
\begin{abstract}
In this article, we demonstrate that the novel general theory of relativity named `Entangled Relativity' is more economical than General Relativity in terms of universal dimensionful constants and units when both theories are considered through a path integral formulation. The sole parameter of Entangled Relativity is a quantum of energy squared. However, in order to recover standard Quantum Field Theory when gravity is neglected in the path integral, we show that this quantum of energy corresponds to the reduced Planck energy. But this result also implies that Planck's quantum of action $\hbar$ and Newton's constant $G$ are not fixed constants in this framework but vary proportionally to a gravitational scalar degree-of-freedom, akin to typical scalar-tensor and $f(R)$ theories. In particular, it is derived that $\hbar$ is proportional to $G$ in this framework. This establishes an explicit connection between the quantum and gravitational realms. Given the absence of any free theoretical parameter in the theory, we evaluate the level of variation of $\hbar$ and $G$ in the solar system and for neutron stars. We argue that this type of quantitative predictions might be probed observationally in the future, although their amplitudes are extremely small. 
\end{abstract}
\maketitle
\section{introduction}
\label{sec:intro}
A major challenge in modern elementary physics is to understand quantum gravity. For decades, it has been asserted that General Relativity and Quantum Field Theory are incompatible, suggesting that merging the two frameworks necessarily leads to a meaningless theory \cite{donoghue:2017sp}. However, as of today, there is absolutely no proof that this is indeed the case. Firstly, at the perturbative level, Quantum General Relativity is perfectly coherent as an Effective Field Theory \cite{donoghue:2017sp,burgess:2004lr,donoghue:2022ax}, enabling the computation of unambiguous quantum corrections to classical phenomenology within this framework. More importantly, theoretical evidence from different lines of research now suggests that non-perturbative Quantum General Relativity might be renormalizable, despite being perturbatively non-renormalizable. This evidence notably comes from the Asymptotic Safety \cite{niedermaier:2006lr,nink:2013sc} and the Causal Dynamical Triangulation \cite{loll:2020cq} programs, which employ different theoretical techniques to explore the potential non-perturbative renormalizability of Quantum General Relativity. Remarkable outcomes from both programs include predictions of a particle physics landscape compatible with an asymptotically safe Quantum General Relativity within the Asymptotic Safety framework \cite{eichhorn:2018fa,dupuis:2021pr,debrito:2022jh}, notably the prediction of the Higgs mass before it was measured \cite{shaposhnikov:2010pl}, and the emergence of a 4-dimensional quantum universe (with a positive renormalized cosmological constant) from first principles in the framework of Causal Dynamical Triangulation \cite{ambjorn:2005pl,glaser:2017cr}. Nevertheless, these approaches have their own open questions and challenges \cite{bonanno:2020fp,loll:2020cq}.

In what follows, we do not argue that Quantum General Relativity has an issue per se, because, to date, no one actually knows \cite{woodard:2023ep}; instead, we propose another potential path toward quantum gravity, based on a novel general theory of relativity that is more economical than General Relativity, while it possesses both General Relativity and standard Quantum Field Theory as predictable limits of the theory. 
Moreover, as we will see, this theory precludes the definition of the Planck units of time and length. Hence, given the central role of Planck time and length in the Quantum Gravity programs to date \cite{kiefer:2012bk}, we argue that Entangled Relativity offers a qualitative departure from other approaches explored thus far.

Indeed, almost ten years ago, an alternative general theory of relativity was proposed, but it was considered a curiosity due to its unusual non-linear Lagrangian density \cite{ludwig:2015pl}. It has recently been named `Entangled Relativity' in \cite{arruga:2021pr}, not because it is related to `quantum entanglement' a priori, but because matter and gravity cannot be treated separately within this framework. Indeed, Entangled Relativity is a general theory of relativity that requires the existence of matter to even be defined, thereby realizing Einstein's original idea that a satisfying theory of relativity should not allow for the existence of vacuum solutions \cite{einstein:1917co,einstein:1918an,einstein:1918sp,einstein:1921bk,pais:1982bk,hoefer:1995cf,minazzoli:2024pn}. Indeed, vacuum solutions (in the classical sense\footnote{That is, in the total absence of matter fields.}) imply that inertia---which is defined from the metric tensor in a relativistic theory---could be defined in the total absence of matter, which would \textit{de facto} violate the \textit{principle of relativity of inertia} \cite{einstein:1917co,einstein:1918an,einstein:1918sp,einstein:1921bk,pais:1982bk,hoefer:1995cf,minazzoli:2024pn} that Einstein named \textit{Mach's principle} in \cite{einstein:1918an}. Despite its very unusual non-linear action---see Eq. (\ref{eq:ERPI}) below---Entangled Relativity has been shown to possess General Relativity as a limit in fairly generic (classical) situations \cite{ludwig:2015pl,arruga:2021pr,arruga:2021ep,minazzoli:2021ej,minazzoli:2021cq,wavasseur:2025gg,minazzoli:2025ep}, indicating that, at least up to further scrutiny, the theory may be viable from an observational standpoint. 

However, it was soon realized that the only parameter of the theory was a quantum parameter, as it does not appear in the field equations \cite{minazzoli:2018pr}. In the present paper, we formulate the theory through its path integral because this approach allows one to explicitly identify this parameter by requiring the theory to be consistent with standard Quantum Field Theory on `flat spacetime'.\footnote{For the author, a `flat spacetime' is only a somewhat useful approximation for scales at which gravity can be neglected, but apart from that, it is not realized anywhere in the universe---as evidenced observationally with the acceleration of the expansion of the universe, and theoretically with the quantum vacuum. Consequently, if considered exactly, a `flat spacetime' is not a \textit{physical} assumption.} 

\section{Formulation and field equations}
\label{sec:formulation}

The path integral formulation of Entangled Relativity reads as follows
\be \label{eq:PI}
Z = \int [\D g]  \prod_j [\D f_j] \exp(i\Theta),
\ee
where the quantum phase is given by
\be
\Theta = -\frac{1}{2 \epsilon^2} \int d^4_g x \frac{\L^2_m(f,g)}{R(g)}, \label{eq:ERPI}
\ee
and where $\int [\D]$ relates to the sum over all possible (non-redundant) field configurations, $R$ is the usual Ricci scalar that is constructed upon the metric tensor $g$, $ \mathrm{d}^4_g x := \sqrt{-|g|}  \mathrm{d}^4 x$ is the spacetime volume element, with $|g|$ the metric $g$'s determinant, and $\L_m$ is the Lagrangian density of matter fields $f$---such as gauge bosons, elementary fermions and the Higgs---which could be the current \textit{standard model of particle physics} Lagrangian density, but most likely a completion of it. It also depends on the metric tensor, a priori through to the usual \textit{comma-goes-to-semicolon rule} \cite{MTW} in order to recover General Relativity in some limit.\footnote{Strictly speaking, this condition is only necessary in the `General Relativity limit' of the theory, but could perhaps be relaxed in general, as long as it then emerges in the required limit.} Let us note that, like General Relativity, Entangled Relativity does not specify what $\Lm$ should be. Moreover, just like General Relativity, Entangled Relativity is a specific $f(R,\Lm)$ theory \cite{harko:2010ep,harko:2014ga,Harko_Lobo_2018}. Given that the dimension of the term in the integral in Eq. (\ref{eq:ERPI}) is an energy squared, the only parameter of the theory is a quantum of energy squared $\epsilon^2$. This means, in particular, that Planck's quantum of action $\hbar$ is not a fundamental constant in this framework, nor is Newton's constant $G$, since they do not appear in the formulation of the theory.

In order to evaluate the limit at which gravity can be neglected, one first need to understand what gravity is in this framework. We do not have the pretension to evaluate the path integral Eq. (\ref{eq:PI}) in this paper, but we can already take advantage of some lessons about classical gravity that we can learn from the study of the paths with stationary phases $\delta \Theta = 0$. As we will see, this alone enables the evaluation of the quantum of energy squared, $\epsilon^2$. Those paths corresponds to the following field equations  \cite{ludwig:2015pl}
\be \label{eq:metric}
G_{\mu \nu} = \kappa T_{\mu \nu} + f_R^{-1} \left[\nabla_\mu \nabla_\nu - g_{\mu \nu} \Box \right] f_R,
\ee
with
\bea
\kappa = - \frac{R}{\Lm},\qquad
f_R = \frac{1}{2 \epsilon^2} \frac{\Lm^2}{R^2} = \frac{1}{2 \epsilon^2 \kappa^2},\label{eq:f_Rkappa}
\eea
with the following stress-energy tensor
\be
T_{\mu \nu} := -\frac{2}{\sqrt{-g}} \frac{\delta\left(\sqrt{-g} \mathcal{L}_{m}\right)}{\delta g^{\mu \nu}},
\ee
which is not classically conserved since
\be
\nabla_{\sigma}\left(\frac{\mathcal{L}_{m}}{R} T^{\alpha \sigma}\right)=\mathcal{L}_{m} \nabla^{\alpha}\left(\frac{\mathcal{L}_{m}}{R}\right). \label{eq:noconsfR}
\ee
The matter field equation, for any tensorial matter field $\chi$, gets modified due to the non-linear coupling between matter and curvature as follows
\bea
\frac{\partial \mathcal{L}_{m}}{\partial \chi}&-&\frac{1}{\sqrt{-|g|}} \partial_{\sigma}\left(\frac{\partial \sqrt{-|g|} \mathcal{L}_{m}}{\partial\left(\partial_{\sigma} \chi\right)}\right) \nonumber\\
&=&\frac{\partial \mathcal{L}_{m}}{\partial\left(\partial_{\sigma} \chi\right)} \frac{R}{\mathcal{L}_{m}} \partial_{\sigma}\left(\frac{\mathcal{L}_{m}}{R}\right). \label{eq:ERmatter}
\eea

\section{Decoupling}
It has already been demonstrated that these equations lead to classical phenomenology very similar to, or even indistinguishable from, that of General Relativity in many cases \cite{ludwig:2015pl,arruga:2021pr,arruga:2021ep,minazzoli:2021ej,minazzoli:2021cq,minazzoli:2025ep,wavasseur:2025gg}. This similarity results from the \textit{intrinsic decoupling} originally identified in scalar-tensor theories \cite{minazzoli:2013pr,minazzoli:2014pr,minazzoli:2014pl}. Specifically, as is common in $f(R)$ theories, the trace of the metric field equation produces the differential equation for the extra scalar degree-of-freedom, 
$f_R$, which is given by:
\begin{equation}
3 f_R^{-1}\Box f_R = \kappa \left(T - \Lm\right). \label{eq:sceq}
\end{equation}
Therefore, whenever $\Lm= T$ on-shell, the extra degree-of-freedom ($f_R\propto \kappa^{-2}$) is not sourced and becomes constant in many cases, allowing one to recover General Relativity, minimally coupled to matter and without a cosmological constant, with very good accuracy \cite{ludwig:2015pl,arruga:2021pr,arruga:2021ep,minazzoli:2021ej,minazzoli:2021cq,minazzoli:2025ep,wavasseur:2025gg}. It is worth noting that $\Lm = T$ is a valid assumption for a universe composed almost entirely of dust and electromagnetic radiation, which closely approximates the current content of our universe. Indeed, assuming that \textit{dark} and \textit{baryonic} matter can be modeled by a dust field, one has $\Lm = -c^2 \sum_i m_i \delta^{(3)}(\vec{x} - \vec{x}_i(t)) / (\sqrt{-g} u^0) = T$, where $m_i$ represents the conserved mass of dust particles along their geodesics \cite{minazzoli:2013pd}; whereas, for electromagnetic radiation, $\Lm \propto E^2 - B^2 = 0 = T$. See also \cite{avelino:2022pr,pinto:2025ar} and references therein for broader arguments implying $\Lm =T$ on-shell.

Let us stress that the whole field equations are well-behaved at the limits $R \rightarrow 0$ and $\Lm \rightarrow 0$, even though it may not be apparent at first glance. Indeed, the behavior of the ratio between $R$ and $\Lm$ is dictated by the entire field equations, and in particular by Eq. (\ref{eq:sceq}), just as the ratio between $R$ and $T$ is constrained by the trace of Einstein's equation in General Relativity. This is exemplified in the spherically and rotating charged black-hole solutions found in \cite{minazzoli:2021ej,wavasseur:2025gg}, which are such that $(\Lm,R) \propto Q^2$, where $Q$ is the charge of the black hole. As a consequence, the ratio between $R$ and $\Lm$, or $\kappa$, turns out to tend to a constant in the $\Lm \rightarrow 0$ limit, which also corresponds to the $R\rightarrow 0$ limit. 
Let us emphasize that when the ratio between $R$ and $\Lm$ becomes constant, one exactly recovers General Relativity minimally coupled to matter fields. Thus, General Relativity emerges as a limit of Entangled Relativity in the regime of weak matter field density. 
This is also exemplified by the solutions for a spherically neutral black hole immersed in a uniform electric or magnetic background, as found in \cite{minazzoli:2025ep}. These solutions tend to the Schwarzschild black hole of General Relativity when the background electric or magnetic field decreases.

Interrestingly, the whole set of Eqs. (\ref{eq:metric}-\ref{eq:sceq}) can be recovered by this alternative Einstein-dilaton phase instead \cite{ludwig:2015pl}
\be \label{eq:ER2QP}
\Theta_{Ed} = \frac{1}{\epsilon^2} \int d^4_g x \frac{1}{\kappa}\left(\frac{R(g)}{2 \kappa} + \Lm(f,g) \right),
\ee
provided that $\L_m \neq \emptyset$, and where $\kappa$ is a dimensionful scalar-field, whose on-shell value matches the definition in Eq. (\ref{eq:f_Rkappa}). This is similar to the usual equivalence between $f(R)$ and Scalar-Tensor theories \cite{capozziello:2015sc}. 
Eq. (\ref{eq:ER2QP}) corresponds to a special case of the theories studied in \cite{minazzoli:2013pr,minazzoli:2014pr}, which are such that $\kappa$ is indeed a weakly sourced gravitational field due to the \textit{intrinsic decoupling} mentioned above. As a consequence, $\kappa$ varies even less than the spacetime metric $g_{\mu \nu}$. In the Solar System, for instance, the metric's perturbation is of order $\mathcal{O}(c^{-2})$, whereas $\kappa$'s perturbation is of order $\mathcal{O}(c^{-4})$, as shown in \cite{minazzoli:2013pr}. The scalar field's perturbation remains smaller than the metric's perturbation, even for neutron stars \cite{arruga:2021pr}, which are the densest objects in the universe that are not hidden behind an event horizon.

\section{Standard particle physics}
\label{sec:particles}

As a consequence, for any quantum phenomenon where gravity can be neglected, the path integral in Eqs. (\ref{eq:PI}-\ref{eq:ERPI}) can be approximated by
\be  \label{eq:ERPISQFT}
Z\approx  \int \prod_j [\D f_j] \exp \left[\frac{i}{\kappa \epsilon^2} \int d^4 x  \L_m(f) \right].
\ee
Therefore, to recover the standard Quantum Field Theory in situations where gravity is negligible, one must ensure that in the limit corresponding to Eq. (\ref{eq:ERPISQFT}), one has
\be \label{eq:id}
\kappa \epsilon^2 = \hbar c.
\ee
This allows one to identify the only free parameter of the theory in Eq. (\ref{eq:ERPI}), $\epsilon^2$, as the squared reduced Planck energy. This is akin to determining the value of the coupling constant $\kappa$ in General Relativity, where $\kappa$ in General Relativity must be chosen so that General Relativity reproduces Newtonian physics in the Newtonian limit. 

\section{Discussion}

In Entangled Relativity, the value of $\kappa$ is determined by its cosmic evolution and by its specific value when it began to stabilize at the onset of the cosmic matter era. For instance, assuming a Friedmann-Lemaître-Robertson-Walker metric with a universe filled with dust and electromagnetic radiation, Eq. (\ref{eq:sceq}) simplifies to $\ddot{f}_R + 3H\dot{f}_R = 0$, with $\kappa^2 \propto f_R^{-1}$ from Eq. (\ref{eq:f_Rkappa}), and where $H$ is the Hubble parameter, leading to $f_R$ (hence $\kappa$) quickly stabilizing ($\dot{f}_R \propto \exp[-3\int H dt]$) close to the value it held during a previous cosmic era.

Eq. (\ref{eq:id}) suggests that the same applies to the value of $\hbar$. Given that $\hbar$ does not appear in Eq. (\ref{eq:ERPI}), it should have been apparent from the outset that $\hbar$ could not be a fundamental constant in Entangled Relativity. Eq. (\ref{eq:id}) indicates that $\hbar$ is an emergent constant, whose constancy is only accurate in the limit where gravity can be entirely neglected. It is important to emphasize that this is not in contradiction with standard physics, as standard Quantum Field Theory, particularly the Standard Model of particle physics, entirely omits gravity from its framework. In fact, in Entangled Relativity, the concept of a \textit{quantum of action} is only pertinent in the semi-classical limit of the theory, where gravity can be treated as a classical background field. At the non-perturbative quantum gravity level, the notion of a \textit{quantum of action} does not exist in Entangled Relativity.\footnote{See Sec. \ref{app:massive} for a discussion on `massive' matter fields.}

This brings us to another significant aspect of Entangled Relativity: the theory lacks sufficient dimensionful universal constants to define Planck's units of time and space. Indeed, the only two dimensionful constants present are the energy squared, $\epsilon^2$, and the causal structure constant, $c$. Considering the pivotal role of the Planck time and length in existing approaches to Quantum Gravity \cite{kiefer:2012bk}, this suggests that Quantum Entangled Relativity could exhibit qualitatively distinct behavior from other approaches in the non-perturbative regime.

Let us indeed note that in Eq. (\ref{eq:f_Rkappa}), one finds $\epsilon^2 \kappa^2 = \kappa \hbar c = \ell_P^2$, where $\ell_P$ represents the reduced Planck length. It is interesting that the new gravitational scalar degree-of-freedom in Entangled Relativity, which arises from the non-linearity of the Lagrangian density, is proportional to the inverse of the squared Planck length, $f_R \propto \ell_P^{-2}$. This illustrates the fact that in Entangled Relativity, the Planck length ($\ell_P$) and time ($\ell_P/c$) are not constants. The only constant is the reduced Planck energy squared, $\epsilon^2$.

Another important lesson from Eq. (\ref{eq:id}) is that, in Entangled Relativity, the weak gravity limit, $\kappa \rightarrow 0$, effectively corresponds to the classical limit, $\hbar \rightarrow 0$. This demonstrates an explicit connection between the quantum and gravitational realms within Entangled Relativity, which would therefore offer a coherent and simplified perspective on elementary physics. Indeed, Eq. (\ref{eq:ERPI}) is simply a non-linear and more economical reformulation of General Relativity.

However, Eq. (\ref{eq:id}) reveals something more profound about Quantum Mechanics and Quantum Field Theory: the procedure of \textit{canonical quantization} should be valid only when gravity can be neglected. Indeed, \textit{canonical quantization} depends on the existence of a constant quantum of action to elevate classical variables (c-numbers) to operators (q-numbers) through Dirac's procedure \cite{birrell:1984bk,wald:1994bk,peskin:1995bk,gambini:2011bk,kiefer:2012bk}. Consequently, since a quantum of action is not a fundamental constant in Entangled Relativity, there's no basis to expect that \textit{canonical quantization} will yield correct results within this framework when gravity cannot be ignored. Actually, Heisenberg's uncertainty principle is also a priori only valid at the limit of the theory where $\kappa$ is constant. But the fact that canonical quantization does not necessarily depict the mathematics underlying nature at a fundamental level is not inconsistent a priori. Indeed, it is possible that the path integral approach in Eq. (\ref{eq:PI}) is the only viable method when dealing with gravity, and that the two approaches are equivalent when gravity is neglected only. Besides, this observation does not challenge established physics, as Quantum Field Theory has been verified experimentally only in conditions where $\hbar$'s variation is negligible---see the numerical evaluations of the variation of $\hbar$ in Sec. \ref{sec:eval}. 

\section{Fields with finite range}
\label{app:massive}

It might be argued that $\hbar$ explicitly appears in the matter Lagrangian $\Lm$ of massive fields in the Standard Model of particle physics. However, fundamentally, what one calls `massive fields' are just `fields with finite range', specified by their (reduced) Compton wavelength $\lambda_C$. This is because any spacetime derivative ($\partial/\partial x^\alpha$) in the kinetic term of massive fields in the matter Lagrangian has to be compensated by a constant with the dimension of length$^{-1}$ in the potential term. 
The reason why $\hbar$ appears in the Lagrangian of standard particle physics is precisely because it is assumed from the outset that $\hbar$ is constant, allowing one to convert the Compton wavelength into a mass scale as $\lambda_C^{-1} = mc/\hbar$. But if $\hbar$ is not a fundamental constant, then one is no longer allowed to do so, and everything has to be kept consistent in terms of dimensions. As a consequence, only the Compton wavelength $\lambda_C$ appears in the definition of fields with finite range when $\hbar$ is not assumed to be constant. 
For instance, the quantum phase of a Dirac field with finite range simply reads
\be \label{eq:thdirac1}
\Theta_{Dirac} = \int d^4x ~\bar \psi(i \slashed{D} - \lambda_C^{-1}) \psi,
\ee
in both standard physics and Entangled Relativity when gravity can entirely be neglected---see Sec. \ref{sec:particles}. Obviously, $\hbar$ plays no role in the definition of a Dirac field with finite range. Similarly, any field with finite range---such as the Higgs field---must involve in its formulation the Compton wavelength that characterizes its finite range. This is imposed by purely dimensional considerations.\\

However, the Lagrangian of matter fields appearing in Eq. (\ref{eq:ERPI}) must have the dimension of an energy density.\\ 

Assuming that $\lambda_C$ is not a fundamental constant---but rather comes from the Englert-Brout-Higgs-Guralnik-Hagen-Kibble mechanism and its dimensionless Yukawa couplings \cite{kibble:2009sc,atlas:2012pl,cms:2012pl,bass:2021na}---$c$, $\epsilon$ and the elementary charge $e$ are the only available dimensionful constants for a Dirac field in Entangled Relativity---see Sec. \ref{sec:units}. Therefore---recalling that $[e^2] = M L^{3} T^{-2}=[\hbar c]$, see Sec. \ref{sec:units}---its Lagrangian must be:
\be
\L_{Dirac} = e^2 ~\bar \Psi(i \slashed{D} - \lambda_C^{-1}) \Psi.
\ee

Using Eq. (\ref{eq:ER2QP}), the resulting quantum phase reads
\be
\Theta_{Dirac} = \int d^4x \frac{e^2}{\epsilon^2 \kappa} \bar \Psi(i \slashed{D} - \lambda_C^{-1}) \Psi,
\ee
which---when gravity can entirely be neglected, such that $\kappa$'s variation can be neglected---can be identified with Eq. (\ref{eq:thdirac1}) with the following field redefinition 
$\psi = \sqrt{e^2/(\epsilon^2 \kappa)} ~\Psi=\sqrt{4\pi \alpha} ~\Psi$, with $\alpha:=e^2/(4\pi \epsilon^2 \kappa)$ being the \textit{non-constant} fine-structure \textit{parameter} in Entangled Relativity.
\section{Fundamental natural units}
\label{sec:units}

\textit{Natural units} are based on three dimensionful constants \cite{uzan:2024ar}. Because neither $G$ nor $\hbar$ are constants in the framework of Entangled Relativity, one has to define new \textit{fundamental natural units} that would be defined from the actual fundamental constants of the theory. Although the status of fundamental constants can evolve over time depending on our understanding of the laws of nature \cite{uzan:2024ar}, we will assume here that the elementary charge $e$ is a fundamental constant and that it has the dimension $[e] = M^{1/2} L^{3/2} T^{-1}$, such that the value of the (here dimensionless) vacuum permittivity $\varepsilon_0$ is one.\footnote{It is argued in \cite{uzan:2024ar} that the electric charge is not universal as it corresponds to the coupling strength of the electromagnetic interaction. However, let us note that all elementary fermions but the neutrinos are charged through this elementary unit in the standard model of particle physics---or thirds of it---such that it appears in the kinetic term of all the interactions of the current standard model (SM) of particle physics but those involving neutrinos. But the kinetic terms of neutrinos depends on the weak force coupling constants, which are also related to the charge $e$ through the Weinberg angle \cite{donoghue:1992bk}. Thus, even in neutrino physics, the unit $e$ appears in their kinetic terms indirectly, such that the unit $e$ effectively appears in all sectors of the SM. Moreover, in a Grand Unified Theory (GUT) of particle physics, there would be only one fundamental charge unit governing all interactions at high energy, which later differentiates into the observed couplings of the SM at lower energies. In any case, only one unit of charge is necessary in physics, as all low-energy coupling constants can be expressed in terms of this unit, scaled by dimensionless parameters that should be provided by the higher energy theory. In this sense, a charge unit $e$ can be considered to be universal.} From there, and from the two dimensionful constants $\epsilon$ and $c$, one can define new \textit{fundamental natural units} of length and time that read
\bea \label{eq:unitl}
&&\mathit{l}_N := \frac{e^2}{\epsilon} = 7.43 \times 10^{-36}~ m,\\
&&\mathit{t}_N := \frac{\mathit{l}_N}{c} = 2.48 \times 10^{-44}~ s,
\eea
where $e = 5.38 \times 10^{-14}~kg^{1/2} m^{3/2} s^{-1}$.\footnote{The unification scale, according to the Standard Model of particle physics seems to be such that $\alpha^{-1} \in [40,47]$ \cite{dias:2004pr}, leading to natural units $\mathit{l}_N  \in [2.17,2.55]\times 10^{-35} m$ and $\mathit{t}_N \in [7.25,8.49]\times 10^{-44}s$---see Sec. \ref{app:massive} for the definition of $\alpha$ in Entangled Relativity. The actual value will be provided by the actual Grand Unified Theory---e.g. $\alpha^{-1} = 38$ in \cite{dias:2004pr}---but it won't impact the discussion overall.} Let us note that unlike Planck length and time, these units are not tied to a notion of elementary units of spacetime a priori. The reduced Planck mass completes the set of \textit{fundamental natural units} in Entangled Relativity as
\be \label{eq:unitm}
\mathit{m}_N := \frac{\epsilon}{c^2} = 4.34 \times 10^{-9}~kg.
\ee

Let us note that there is a degeneracy of \textit{fundamental natural units} in the current standard theory of physics---also known as the Core theory \cite{wilczek:2016bk}---because there are too many fundamental dimensionful constants, which are $e,\hbar,G,c$ and arguably also the cosmological constant. One can notably name the Johnstone-Stoney and the Planck units \cite{uzan:2024ar}, as well as the units defined in Eqs. (\ref{eq:unitl}-\ref{eq:unitm}). This degeneracy entirely disappears in the framework of Entangled Relativity, since it is based on only three fundamental dimensionful constants, which are $e,\epsilon$ and $c$. Therefore, only remains the \textit{fundamental natural units} defined in Eqs. (\ref{eq:unitl}-\ref{eq:unitm}).
\footnote{The Higgs sector of the SM contains a length scale associated with electroweak symmetry breaking $\ell_H \sim 10^{-18}~m$. However, it is assumed in the last paragraph  that this scale is not a fundamental constant itself, as expected in many beyond-SM models.}

\section{Numerical evaluation of the variation of $\hbar$}
\label{sec:eval}

Eq. (\ref{eq:id}) enables the derivation of the expected numerical amplitude for the variation of $\hbar$ in various contexts. Using the post-Newtonian analysis from \cite{minazzoli:2013pr}, it can be determined that within the Solar System, for example, the anticipated relative numerical variation of $\hbar$ between the surface of the Sun and Earth is
\be
\frac{\Delta \hbar}{\hbar} = \frac{G M^P_\odot}{c^2}\left(\frac{1}{r_\odot}-\frac{1}{r_\oplus} \right) \approx \frac{G M^P_\odot}{r_\odot c^2} \sim 2.4\times 10^{-12}, \label{eq:num}
\ee
where $r_\odot$ and $r_\oplus$ are the position of the surface of the Sun and of the Earth respectively, in heliocentric coordinates
, and where a new type of mass term for a given body $A$, produced solely by pressure, has been defined as follows:
\be \label{eq:Mpdef}
M_A^P := \int_A \frac{P(r)}{c^2}  d^3r.
\ee
The numerical evaluations that led to Eq. (\ref{eq:num}) can be found at \url{https://github.com/ominazzoli/hbar-in-SS}, and rely on the model S \cite{christensen-Dalsgaard:1996sc} for the Sun's pressure.\\

Let us emphasize that Eq. (\ref{eq:num}) is independent of any free theoretical parameters.\footnote{That being said, the numerical evaluation nevertheless relies on the assumption that $\Lm = - \rho$ for the Sun \cite{minazzoli:2013pr}, which corresponds to the on-shell Lagrangian of perfect fluids with conserved rest-mass particles \cite{minazzoli:2012pr,arruga:2021pr}. The on-shell Lagrangian for the matter fields that make up the Sun has yet to be derived from \textit{first principles},  possibly along the lines of what was done in \cite{pinto:2025ar} for a K-monopole toy model.} This was also the case for the predictions of General Relativity, in contrast to most modern alternative gravitational theories \cite{clifton:2012pr,berti:2015cq,joyce:2015pr,fienga:2024lr}. Moreover, let us also stress that the theory is indistinguishable from General Relativity at the post-Newtonian level \cite{minazzoli:2013pr}.\\

The largest variation of $\hbar$ in the observable universe is expected between the surface of a neutron star and a distant observer. Using Eq. (\ref{eq:id}), numerical simulations from \cite{arruga:2021pr,arruga:2021ep} estimate this variation to be at the level of a few percent for the densest neutron stars conceivable. 
Although these simulations did not consider the impact of the variation in $\hbar$ on the neutron star's equation of state, the relatively minor extent of this variation suggests that this approximation was indeed a reasonable starting point, unlikely to significantly affect the estimations in \cite{arruga:2021pr,arruga:2021ep}.
\section{Conclusion}

Entangled Relativity predicts that the quantum of action $\hbar$ is not a fundamental constant of nature but emerges as a constant only in the limit where gravity can be entirely neglected. The potential variation of $\hbar$ is relevant not only to the community interested in gravity but also to a broader range of physicists, as it may impact other aspects of quantum physics, such as quantum entanglement between remote particles in different gravitational fields, or possibly even decoherence. 
Nevertheless, given the minuscule level of variation of $\hbar$ in the solar system evaluated in Sec. \ref{sec:eval}, the predicted variation of $\hbar$ does not impact much how Quantum Mechanics and Quantum Field Theory describe quantum phenomena at the experimental level on Earth. However, it has also been argued in Sec. \ref{sec:eval} that the variation of $\hbar$ could reach the percent level for the most compact objects in the universe, thereby also providing a potential way to check this prediction. Should the variation of $\hbar$ be quantitatively confirmed at the observational or experimental level, it would likely imply that Entangled Relativity is better than General Relativity in order to describe the relativistic laws of physics in general. This would stem not only from a theory that is more economical than General Relativity in terms of fundamental constants and units, see Secs. \ref{sec:formulation} and \ref{sec:units}, but also from a theory that better aligns with the whole set of principles Einstein initially proposed to construct General Relativity \cite{einstein:1917co,einstein:1918an,einstein:1918sp,einstein:1921bk,pais:1982bk,hoefer:1995cf,minazzoli:2024pn}; while effectively reducing to General Relativity in many instances to an extremely good level of accuracy \cite{ludwig:2015pl,arruga:2021pr,arruga:2021ep,minazzoli:2021ej,minazzoli:2021cq,wavasseur:2025gg,minazzoli:2025ep}---see Secs. \ref{sec:intro} and \ref{sec:eval}.

\section*{Acknowledgements}
I would like to extend my gratitude to Alexander Vikman for his insightful question regarding the non-fundamental nature of $\hbar$ in Entangled Relativity, posed during the 56th Rencontres de Moriond. His inquiry prompted me to articulate and refine my initial ideas on the subject. I also wish to express my sincere gratitude to Aurélien Hees, Nelson Christensen, Mairi Sakellariadou, Cliff Burgess, and John Donoghue for their invaluable moral support during the challenging previous peer review process of this paper, which included multiple desk rejections.

\bibliography{ER_QoA}
%
%
%







\end{document}